\documentclass{emulateapj}
\usepackage{rotating,psfig}

\newcommand{\eqn}[1]{equation~(\ref{#1})}
\newcommand{\fig}[1]{Figure~\ref{#1}}
\newcommand{\tab}[1]{Table~\ref{#1}}

\newcommand{\nbzh}{30}
\newcommand{\nbzhnob}{15}
\newcommand{\sersic}{S\'ersic}
\newcommand{\hlim}{25}
\newcommand{\nstar}{75}

\newcommand{\sex}{\texttt{SExtractor}}
\newcommand{\multidrizzle}{\texttt{MultiDrizzle}}
\newcommand{\galfit}{\texttt{GalFit}}

\newcommand{\hst}{{\it HST}}
\newcommand{\vlt}{{\it VLT}}
\newcommand{\sst}{{\it Spitzer}}

\shorttitle{Passively Evolving Galaxies in the WFC3 ERS}
\shortauthors{Ryan Jr. et al.}

\begin{document}

\title{The Size Evolution of Passive Galaxies: Observations from the Wide Field Camera 3 Early Release Science Program\footnote{Based on observations made with the NASA/ESA Hubble Space Telescope, obtained from the Data Archive at the Space Telescope Science Institute, which is operated by the Association of Universities for Research in Astronomy, Inc., under NASA contract NAS 5-26555.}}

\author{R. E. Ryan Jr.\altaffilmark{2}, 
P. J. McCarthy\altaffilmark{3}, 
S. H. Cohen\altaffilmark{4}, 
H. Yan\altaffilmark{5}, 
N. P. Hathi\altaffilmark{6}, 
A. M. Koekemoer\altaffilmark{7}, 
M. J. Rutkowski\altaffilmark{4},
M. R. Mechtley\altaffilmark{4},
R. A. Windhorst\altaffilmark{4}, 
R. W. O'Connell\altaffilmark{8},  
B. Balick\altaffilmark{9}, 
H. E. Bond\altaffilmark{7}, 
H. Bushouse\altaffilmark{7}, 
D. Calzetti\altaffilmark{10}, 
R. M. Crockett\altaffilmark{11}, 
M. Disney\altaffilmark{12}, 
M. A. Dopita\altaffilmark{13}, 
J. A. Frogel\altaffilmark{14}, 
D. N. B. Hall\altaffilmark{15}, 
J. A. Holtzman\altaffilmark{16}, 
S. Kaviraj\altaffilmark{11},
R. A. Kimble\altaffilmark{17}, 
J. MacKenty\altaffilmark{7}, 
M. Mutchler\altaffilmark{7}, 
F. Paresce\altaffilmark{18}, 
A. Saha\altaffilmark{19}, 
J. I. Silk\altaffilmark{11}, 
J. Trauger\altaffilmark{20}, 
A. R. Walker\altaffilmark{21}, 
B. C. Whitmore\altaffilmark{7}, 
and E. Young\altaffilmark{22}}

\email{rryan@physics.ucdavis.edu}

\altaffiltext{2}{Physics Department, University of California, Davis, CA 95616}
\altaffiltext{3}{Observatories of the Carnegie Institute of Washington, Pasadena, CA 91101}
\altaffiltext{4}{School of Earth and Space Exploration, Arizona State University, Tempe AZ 85287}
\altaffiltext{5}{Center for Cosmology and Astroparticle Physics, Ohio State University, Columbus, OH 43210}
\altaffiltext{6}{Department of Physics and Astronomy, University of California, Riverside, CA 92521}
\altaffiltext{7}{Space Telescope Science Institute, Baltimore, MD 21218}
\altaffiltext{8}{Department of Astronomy, University of Virginia, Charlottesville, VA 22904}
\altaffiltext{9}{Department of Astronomy, University of Washington, Seattle, WA 98195}
\altaffiltext{10}{Department of Astronomy, University of Massachusetts, Amherst, MA 01003}
\altaffiltext{11}{Department of Physics, University of Oxford, Oxford OX1 3PU, United Kingdom}
\altaffiltext{12}{School of Physics and Astronomy, Cardiff University, Cardiff CF24 3AA, United Kingdom}
\altaffiltext{13}{Research School of Astronomy and Astrophysics, The Australian National University, Weston Creek, ACT 2611, Australia}
\altaffiltext{14}{Association of Universities for Research in Astronomy, Washington, DC 20005}
\altaffiltext{15}{Institute for Astronomy, University of Hawaii, Honolulu, HI 96822}
\altaffiltext{16}{Department of Astronomy, New Mexico State University, Las Cruces, NM 88003}
\altaffiltext{17}{NASA, Goddard Space Flight Center, Greenbelt, MD 20771}
\altaffiltext{18}{Istituto di Astrofisica Spaziale e Fisicia Cosmica, INAF, via Gobetti 101, 40129 Bologna, Italy}
\altaffiltext{19}{National Optical Astronomy Observatories, Tucson, AZ 85726}
\altaffiltext{20}{NASA, Jet Propulsion Laboratory, Pasadena, CA 91109}
\altaffiltext{21}{Cerro Tololo Inter-American Observatory, La Serena, Chile}
\altaffiltext{22}{NASA, Ames Research Center, Moffett Field, CA 94035}

\begin{abstract}

We  present  results  on  the  size evolution  of  passively  evolving
galaxies  at $1\!\lesssim\!z\!\lesssim\!2$ drawn  from the  Wide Field
Camera 3  Early Release Science  program.  Our sample  was constructed
using  an  analog to  the  passive  $BzK$  selection criterion,  which
isolates  galaxies  with  little  or  no on-going  star  formation  at
$z\!\gtrsim\!1.5$.      We     identify     \nbzh\     galaxies     in
$\sim\!40$~arcmin$^2$ to  $H\!<\!25$~mag.  We supplement spectroscopic
redshifts  from the literature  with photometric  redshifts determined
from  the   15-band  photometry  from   $0.22-8~\mu$m.   We  determine
effective radii from \sersic\ profile fits to the $H$-band image using
an empirical PSF.  We find that size evolution is a strong function of
stellar mass,  with the most  massive ($M_*\!\sim\!10^{11}~M_{\odot}$)
galaxies undergoing the most  rapid evolution from $z\!\sim\!2$ to the
present.  Parameterizing  the size evolution  as $(1+z)^{-\alpha}$, we
find  a  tentative  scaling  between  $\alpha$  and  stellar  mass  of
$\alpha\!\approx\!-1.8+1.4\log(M/10^{9}~M_{\odot})$.     We    briefly
discuss the implications  of this result for our  understanding of the
dynamical evolution of the red galaxies.

\end{abstract}


\section{Introduction}

Understanding  the red, passively  evolving, galaxies  at intermediate
redshifts  ($z\!\sim\!2$)  is one  of  the  outstanding challenges  of
galaxy  evolution studies.   Early expectations  of a  high luminosity
phase associated with  the formation of spheroids on  a free-fall time
scale  \citep{els} have long  given way  to a  framework in  which the
spheroids  are  assembled  over  an  extended  period  of  time.   The
identification of fairly high space densities of massive and passively
evolving  galaxies at  $z\!\gtrsim\!1.5$ \citep[e.g.][]{glaze04,cim04}
revealed  weaknesses  in  the  early  semi-analytic  models  of  their
formation.  Large-scale surveys, both locally and out to $z\!\sim\!1$,
provide a fairly clear view of  the evolution in the number density of
red sequence galaxies  and the global stellar mass  density in passive
systems \citep[e.g.][]{bell06,faber07,brown03,brown07}.

Deep       spectroscopic      studies       of       red      galaxies
\citep[e.g.][]{mcc04,cim06,cim08}  estimated  stellar ages  consistent
with   early   formation   redshifts   ($z_{\rm   form}\!\gtrsim\!4$),
potentially at odds  with the rapid evolution in  stellar mass density
at these epochs \citep[e.g.][]{rudnick}.  It is now clear that the red
sequence was  in place  at fairly early  epochs \citep{bell04,bremer},
but   evolved  strongly   in  the   $1\!\lesssim\!z\!\lesssim\!3$  era
\citep[e.g.][]{demar}   and   likely   since  $z\!\sim\!1$   as   well
\citep[e.g.][]{faber07}.   The  emerging   picture  in  which  massive
galaxies are  assembled via mergers  followed by a rapid  quenching of
star formation  addresses many  of the salient  properties of  the red
sequence galaxies \citep{faber07}.

In light  of these results it  was quite surprising  that high spatial
resolution   studies  showed   that  the   passive  red   galaxies  at
intermediate redshifts  are systematically smaller  than their likely
present-day counterparts \citep[e.g.][]{dad05,tru06}.  Observations of
$1\!\lesssim\!z\!\lesssim\!2$  passive galaxies  with  either Advanced
Camera  of Surveys  (ACS)  or Near  Infrared  Camera and  Multi-Object
Spectrometer (NICMOS) on the  {\it Hubble Space Telescope} (\hst) gave
sizes  that are  a factor  of $\sim\!2$  smaller than  equal  mass red
galaxies today  \citep[e.g.][]{long07,cim08,dam09}, while observations
of  red  galaxies  at   $z\!\gtrsim\!2$  suggest  even  more  dramatic
evolution \citep[e.g.][]{vd08,sar10}.

The  primary  interest  in  determining the  characteristic  sizes  of
passive galaxies  is the insight  that it provides into  the dynamical
state  of the  system.  As  one  projection of  the fundamental  plane
\citep{dd,dres}, the Kormendy relation \citep{korm77} provides a probe
of the dynamics of hot  stellar systems.  The Kormendy diagram for the
$z\!\gtrsim\!1$  red sequence  galaxies is  nearly as  tight as  it is
locally,  but   is  offset  in   both  surface  brightness   and  size
\citep[e.g.][]{dam09}, even when the  effects of passive fading of the
stellar populations is taken into account.  The compact sizes and high
surface brightness  of the red  galaxies implies stellar  densities in
the  centers  of  these galaxies  that  are  two  to three  orders  of
magnitude   higher    than   present   day    massive   red   galaxies
\citep[e.g.][]{vd08,dam09}.

The challenge  in understanding these results stems  from the apparent
conflict between the  requirement for a factor of  $\sim\!3$ growth in
size during an epoch in which the stellar masses, population ages, and
overall morphologies show little or  no evolution.  A number of models
are under  active discussion, but  most involve large numbers  of late
stage  minor   mergers  that  grow   the  galaxies  in   size  without
contributing  large  amounts  of  mass.   While  it  is  difficult  to
conclusively rule out various measurement biases which might give rise
to  this  apparent   trend  (e.g.~underestimated  effective  radii  or
overestimated  masses),  most   interpretations  are  astrophysical  in
origin:  (major  or  minor)  merging  occurring  at  different  phases
\citep[e.g.][]{naab06}, or  expansion due to a  significant mass loss,
either by  an active galactic  nucleus \citep{fan08} or  stellar winds
\citep{dam09}.  Recently  \citet{hop10} present a  semi-analytic model
based on high-resolution hydrodynamic simulations \citep{cox06}, which
incorporates  several astrophysical  and observational  mechanisms for
this  trend, and conclude  that for  galaxies with  a stellar  mass of
$M_*\!\geq\!10^{11}~M_{\odot}$,  the late-stage  minor merging  is the
dominate mechanism,  accounting for $\sim\!50\%$ of  the apparent size
evolution.  Certainly  many of these high-redshift systems  do in fact
show merger-like features \citep[e.g.][]{vd05,vd10} or have (multiple)
companions \citep[e.g.][]{bell06}.

The  merging  scenario   implies  that  some  non-negligible  fraction
\citep[$\lesssim\!10\%$;][]{hop09}  of compact galaxies  should remain
to   $z\!\sim\!0$.     However   \citet{tru09}   find    only   0.03\%
($=48/152,083$)  galaxies  at  $z\!\sim\!0.2$ have  stellar  densities
comparable to these high redshift galaxies, arguing against the merger
scenario.  To further confound the issue, these galaxies are generally
young ($\sim\!3$~Gyr old), giving an approximate formation redshift of
$z_{\rm form}\!\sim\!0.3$, suggesting that  they are not the relics of
the early  universe, but formed from gas-rich,  recently merged disks.
Conversely, \citet{sar10}  find that $\sim\!62\%$  ($=21/34$) of their
galaxies  at $1\!<\!z_{\rm  spec}\!<\!2$ are  within 1$\sigma$  of the
local $R_e-M_*$ relation \citep{shen03}.   Based on this finding, they
suggest  that   the  compact  high  redshift  galaxies   are  not  the
progenitors of large field  galaxies, but of compact brightest cluster
galaxies.  The absence of  compact early-type galaxies at $z\!\sim\!0$
seems to suggest  that the simple picture of  galaxy merging with some
mass-loss expansion  is not completely  correct, in stark  contrast to
the simulated results \citep[e.g.][]{naab09,hop10}.

This paper  is organized as  follows. In \S~\ref{obs} we  describe the
observations  and ancillary  data,  in \S~\ref{sample}  we define  our
sample, in \S~\ref{analysis} we detail our stellar population modeling
and  morphological  analysis,  in  \S~\ref{sizeevol} we  describe  our
mass-dependent  size  evolution  model,  in \S~\ref{disc}  we  discuss
several key results,  and in \S~\ref{summary} we give  a brief summary
of this work with thoughts  for future surveys.  Throughout this paper
we   assume  a   $\Lambda$CDM   cosmology  with   $\Omega_0\!=\!0.27$,
$\Omega_{\Lambda}\!=\!0.73$,  and  $H_0\!=\!73$~km~s$^{-1}$~Mpc$^{-1}$
\citep{sper03} and  will quote all magnitudes in  the ${\rm AB}_{\nu}$
magnitude system \citep{abmag}.

\section{Observations} \label{obs}

\subsection{The Wide Field Camera 3 Early Release Science}

For  this work  we analyze  the  Early Release  Science (ERS;  PropID:
11359, PI: R.~W.~O'Connell) observations conducted with the Wide-Field
Camera~3 (WFC3) recently installed on the {\it Hubble Space Telescope}
(\hst).  This field covers roughly the northern $40$~arcmin$^2$ of the
Great    Observatories   Origins    Deep    Survey   southern    field
\citep[GOODS-S;][]{giav04}, and leverages the existing Advanced Camera
for  Surveys  (ACS)  optical  data  with  near-ultraviolet  (NUV)  and
near-infrared  (NIR) data  with equivalent  space-based  imaging.  The
main  details regarding the  data collection,  reduction, calibration,
and mosaicking  are presented by \citet{win10}.   We briefly summarize
the observational aspects critical to this work.

The ERS program utilizes the two complementary modes of the WFC3: UVIS
and  IR  channels.  The  data  were  collected  between September  and
November   2009   and  form   a   $2\times4$  and   $2\times5$~mosaic,
respectively. The  final dataset consists  of $\sim\!40$~arcmin$^2$ of
10-band \hst\ imaging covering  2250~\AA\ to 1.6~$\mu$m in wavelength.
The v2.0 GOODS high-level science products had an original pixel scale
of  30~milliarcseconds, and  were $3\times3$~block  summed  to produce
science and weight maps of  equal pixel scale to the WFC3 observations
\citep[see][for a  more detailed  discussion on the  rebinning process
  and its motivation]{win10}.  Therefore, our final \hst\ mosaics have
90~milliarcsecond pixels.  We refer  hereafter to the 10-band WFC3 and
ACS  filter set  in  the  F225W, F275W,  F336W,  F435W, F606W,  F775W,
F850LP, F098M, F125W, and  F160W as $U_1U_2U_3BVi'z'Y_sJH$ to simplify
the bandpass notation.

\subsection{Source Catalogs} \label{sexcat}

We created source catalogs  using \sex\ \citep{bert} with the $H$-band
mosaic  as  the detection  image  and each  of  the  \hst\ mosaics  as
measurement   images.    We   use   the  weight   maps   produced   by
\multidrizzle\ \citep{mdriz}, modified to account for correlated pixel
noise  \citep{d04}.  For  object detection,  we require  a  minimum of
5~connected  pixels  each greater  than  $0.6\sigma$  above the  local
background  and  apply  a  3~pixel  Gaussian  smoothing  filter.   For
deblending,  we   adopt  a   contrast  parameter  of   $10^{-4}$  with
64~sub-thresholds.   For photometry,  we adopt  the \texttt{MAG\_AUTO}
measurements with a  Kron factor of 2.5, and  minimum object radius of
3.5 pixels, which reliably  recover total fluxes to within $\sim\!6$\%
\citep{bert}.  We use the AB zeropoints given by \citet{kali31,kali30}
for the UVIS and IR data, respectively.

\section{Sample Selection} \label{sample}

\begin{figure}
\epsscale{1.20}
\plotone{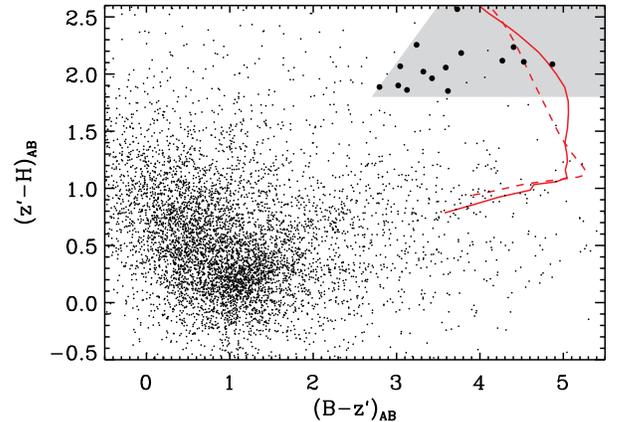}
\caption{The  $BzH$ object  selection.  The  small dots  represent all
  objects detected and  measured by \sex.  The grey  area is the color
  selection    region   defined    by    equations~\ref{bzheqn0}   and
  \ref{bzheqn1}       with        a       typical       color       of
  $\left<(H-K)\right>\!=\!0.7$~mag.    We  identify   \nbzh\  galaxies
  strictly  meeting the $BzH$  color criteria  (shown as  large filled
  circles), however \nbzhnob\ galaxies have $(B-z')$ colors too red to
  be shown  here. There are  several galaxies which formally  meet our
  color criteria but were rejected as  being too faint or too close to
  survey edges  to be studied. The  red lines show the  tracks for the
  \citet[][dashed]{cww80}  and  \citet[][solid]{cb07}  templates.   We
  note, Galactic  stars will span roughly similar  $(B-z')$ colors but
  will  be  $\gtrsim\!0.7$~mag  too  blue  in $(z'-H)$  to  have  been
  misidentified        as       $BzH$        galaxies       \citep[see
    also][]{daddi04}.\label{bzhselect}}
\end{figure}

\begin{figure*}
\centerline{\psfig{file=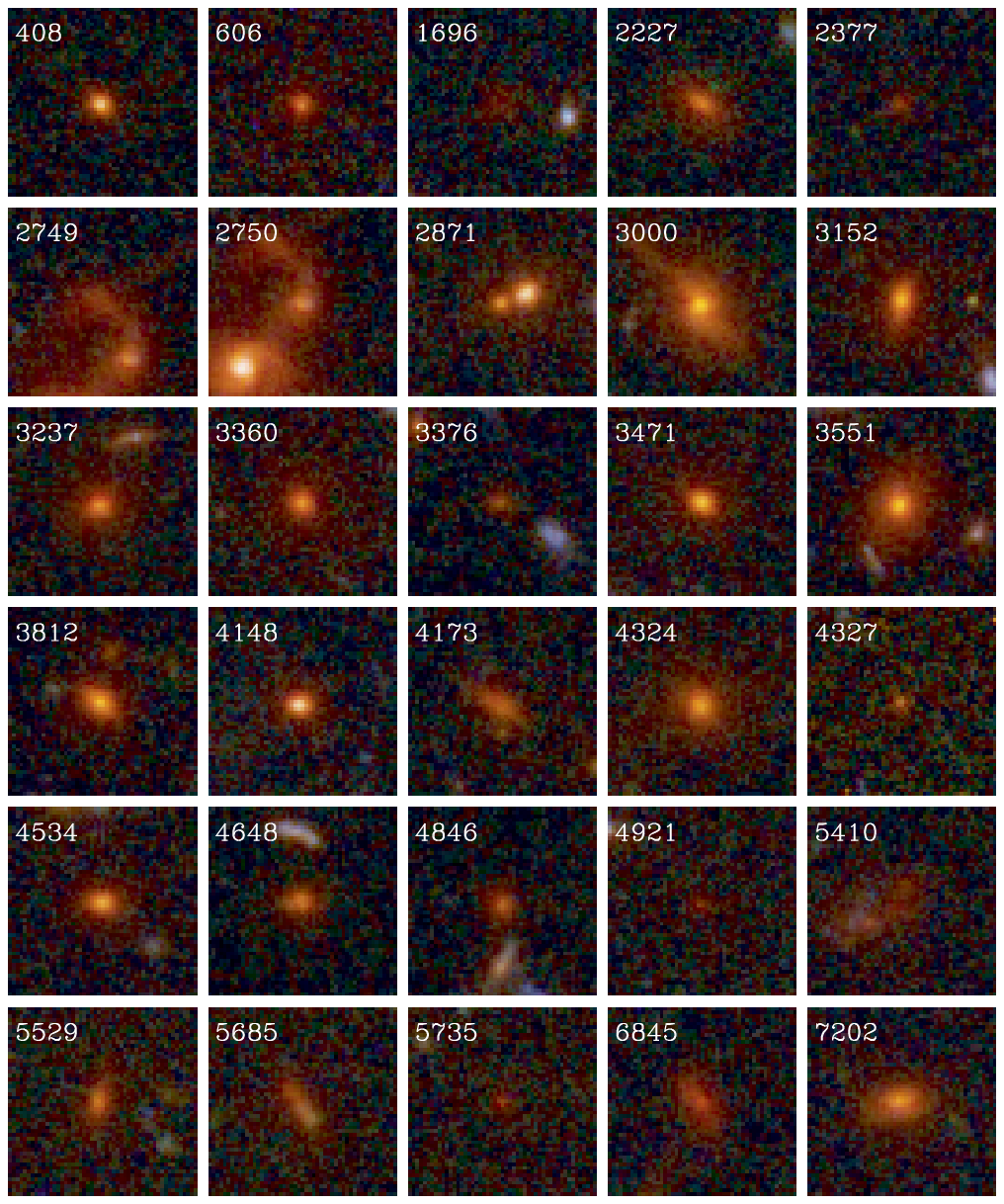,width=5.2truein}}
\caption{$BzH$  color  images.    Each  stamp  is  $\approx\!4\farcs5$
  ($\approx\!38$~kpc at $z\!=\!1.6$) on a  side, has north up and east
  left, and  has a pixel scale of  $0\farcs090$~pix$^{-1}$. All images
  are shown  with the  same color and  logarithmic intensity  scale. We
  find the  object just to  the south of  \#2750 (and \#2749)  is just
  barely  too blue  to  have been  included  in our  sample, which  is
  consistent  with  the  suggestion  that  this  object  has  recently
  ($\sim\!150$~Myr) undergone  an intense  burst of star  formation of
  $500-2000~M_{\odot}$~yr$^{-1}$ \citep{vd10}. \label{colim}}
\end{figure*}

We select  our passively  evolving galaxies using  a variation  on the
standard  p$BzK$  selection   \citep{daddi04},  designed  to  identify
galaxies  at  $z\!\sim\!2$ with  little  on-going  star formation  and
stellar population  ages of $\sim\!1$~Gyr  \citep{dad05}.  However, to
take full  advantage of our high-spatial  resolution and significantly
deeper  ERS  dataset, we  use  our  $H$-band  imaging instead  of  the
ground-based  $K$-band  images.   Therefore  we modify  the  classical
p$BzK$ selection criteria to an equivalent p$BzH$ scheme:
\begin{eqnarray} \label{bzheqn0}
(z'-H)-(B-z')&<&-0.2-\left<(H-K)\right>~\mathrm{mag},\\
(z'-H)&>&2.5-\left<(H-K)\right>~\mathrm{mag}. \label{bzheqn1}
\end{eqnarray}
As  these  p$BzK$  galaxies  have  maximally-old,  passively  evolving
systems, we derive a  typical $\left<(H-K)\right>$ color indicative of
a  stellar  population with  an  instantaneous star-formation  history
formed  at  $z_{\rm  form}\!=\!10$  based  on  the  \citet[][hereafter
  CB07]{cb07}   models    of   $\left<(H-K)\right>\!=\!0.7$~mag.    In
\fig{bzhselect}, we  show the $(B-z')$ and $(z'-H)$  colors with these
criteria illustrated as  a shaded polygon.  To rule  out any potential
image artifacts  associated with the  ERS field edges, we  require the
objects to be in the portion of the $H$-band mosaic which received the
full complement of two orbits  per pointing.  We restrict the $H$-band
magnitude to  a relatively conservative  limit of $H\!\leq\!\hlim$~mag
to ensure  a complete and  reliable sample \citep{win10}.   With these
requirements,  we  identify  a  final  sample of  \nbzh\  galaxies  in
$43.1$~arcmin$^2$.  Of these  galaxies, \nbzhnob~objects have $(B-z')$
colors  redder  than  the  plot  limits and  are  not  represented  in
\fig{bzhselect}.     We   tabulate    the    \nbzh\   candidates    in
\tab{photometry} and show color stamps of each object in \fig{colim}.

\begin{sidewaystable*}
\caption{$BzH$ Sample and \hst\  Photometry}
\label{photometry}
\begin{tabular*}{0.98\textwidth}
  {@{\extracolsep{\fill}}lrrcccccccl}
\hline\hline
\multicolumn{1}{c}{ID} & \multicolumn{1}{c}{RA$^\dagger$} & \multicolumn{1}{c}{Dec$^\dagger$} & \multicolumn{1}{c}{$B$} & \multicolumn{1}{c}{$V$} & \multicolumn{1}{c}{$i'$} & \multicolumn{1}{c}{$z'$} & \multicolumn{1}{c}{$Y_s$} & \multicolumn{1}{c}{$J$} & \multicolumn{1}{c}{$H$} & \multicolumn{1}{c}{Notes}\\
\multicolumn{1}{c}{$ $} & \multicolumn{1}{c}{($^h\;^m\;^s$)} & \multicolumn{1}{c}{($^{\circ}\;'\;"$)} & \multicolumn{1}{c}{(mag)} & \multicolumn{1}{c}{(mag)} & \multicolumn{1}{c}{(mag)} & \multicolumn{1}{c}{(mag)} & \multicolumn{1}{c}{(mag)} & \multicolumn{1}{c}{(mag)} & \multicolumn{1}{c}{(mag)} & \multicolumn{1}{c}{$ $}\\
\hline
  408 & $3\;32\;28.02$ & $-27\;40\;31.2$ & $27.52\pm0.49$ & $26.18\pm0.12$ & $25.18\pm0.08$ & $24.40\pm0.05$ & $23.77\pm0.03$ & $22.87\pm0.01$ & $22.54\pm0.01$ & \nodata \\
  606 & $3\;32\;23.27$ & $-27\;40\;45.8$ & $>\!28.11$     & $27.60\pm0.53$ & $26.12\pm0.23$ & $25.36\pm0.14$ & $25.16\pm0.17$ & $23.66\pm0.03$ & $23.10\pm0.02$ & \nodata \\
 1696 & $3\;32\;42.08$ & $-27\;41\;41.2$ & $>\!27.36$     & $>\!27.29$     & $26.69\pm1.12$ & $26.16\pm0.77$ & $25.47\pm0.32$ & $24.98\pm0.14$ & $24.08\pm0.08$ & \nodata \\
 2227 & $3\;32\;42.34$ & $-27\;42\;04.0$ & $>\!27.55$     & $26.85\pm0.63$ & $24.53\pm0.13$ & $24.50\pm0.14$ & $23.86\pm0.07$ & $22.91\pm0.02$ & $22.32\pm0.01$ & \nodata \\
 2377 & $3\;32\;25.03$ & $-27\;42\;09.6$ & $>\!28.10$     & $27.80\pm0.67$ & $27.68\pm1.04$ & $27.43\pm1.00$ & $26.44\pm0.42$ & $25.42\pm0.11$ & $24.52\pm0.06$ & X-ray ID 226$^\ddagger$ \\
 2749 & $3\;32\;14.92$ & $-27\;42\;21.9$ & $>\!27.23$     & $26.40\pm0.44$ & $24.92\pm0.19$ & $24.64\pm0.19$ & $24.67\pm0.19$ & $23.22\pm0.03$ & $22.69\pm0.03$ & \nodata \\
 2750 & $3\;32\;14.88$ & $-27\;42\;23.3$ & $>\!27.26$     & $25.70\pm0.22$ & $24.56\pm0.14$ & $23.92\pm0.09$ & $23.95\pm0.09$ & $22.46\pm0.02$ & $21.97\pm0.01$ & X-ray ID 145$^\ddagger$ \\
 2871 & $3\;32\;31.09$ & $-27\;42\;26.6$ & $>\!28.23$     & $26.85\pm0.22$ & $25.78\pm0.14$ & $24.91\pm0.08$ & $24.17\pm0.05$ & $23.39\pm0.02$ & $23.07\pm0.02$ & \nodata \\
 3000 & $3\;32\;43.93$ & $-27\;42\;32.4$ & $26.23\pm0.47$ & $25.12\pm0.13$ & $23.61\pm0.06$ & $22.64\pm0.03$ & $21.99\pm0.02$ & $21.07\pm0.00$ & $20.59\pm0.00$ & \nodata \\
 3152 & $3\;32\;26.05$ & $-27\;42\;36.6$ & $>\!27.96$     & $27.21\pm0.44$ & $24.93\pm0.09$ & $23.93\pm0.05$ & $23.31\pm0.02$ & $22.15\pm0.01$ & $21.66\pm0.00$ & \nodata \\
 3237 & $3\;32\;35.91$ & $-27\;42\;40.9$ & $>\!28.04$     & $26.77\pm0.30$ & $25.24\pm0.13$ & $24.60\pm0.08$ & $23.90\pm0.05$ & $22.67\pm0.01$ & $22.06\pm0.01$ & \nodata \\
 3360 & $3\;32\;30.58$ & $-27\;42\;43.4$ & $>\!28.06$     & $27.27\pm0.44$ & $25.90\pm0.22$ & $24.65\pm0.08$ & $24.11\pm0.05$ & $22.78\pm0.01$ & $22.22\pm0.01$ & \nodata \\
 3376 & $3\;32\;35.13$ & $-27\;42\;37.0$ & $>\!28.04$     & $27.95\pm0.74$ & $26.80\pm0.45$ & $26.69\pm0.48$ & $25.40\pm0.16$ & $24.91\pm0.07$ & $24.55\pm0.06$ & \nodata \\
 3471 & $3\;32\;27.94$ & $-27\;42\;45.7$ & $28.29\pm1.34$ & $26.29\pm0.17$ & $24.75\pm0.07$ & $24.02\pm0.04$ & $23.48\pm0.02$ & $22.35\pm0.01$ & $21.91\pm0.00$ & \nodata \\
 3488 & $3\;32\;25.65$ & $-27\;42\;46.8$ & $>\!27.97$     & $>\!28.20$     & $>\!27.63$     & $26.95\pm0.70$ & $26.01\pm0.31$ & $25.38\pm0.12$ & $24.93\pm0.10$ & \nodata \\
 3551 & $3\;32\;36.28$ & $-27\;42\;49.4$ & $27.78\pm1.42$ & $26.05\pm0.24$ & $24.32\pm0.08$ & $23.38\pm0.04$ & $22.78\pm0.02$ & $21.63\pm0.00$ & $21.14\pm0.00$ & \nodata \\
 3812 & $3\;32\;36.66$ & $-27\;42\;58.5$ & $>\!28.02$     & $26.05\pm0.15$ & $24.87\pm0.09$ & $23.75\pm0.04$ & $23.35\pm0.02$ & $22.23\pm0.01$ & $21.74\pm0.00$ & \nodata \\
 4148 & $3\;32\;44.97$ & $-27\;43\;09.1$ & $27.33\pm0.48$ & $25.00\pm0.05$ & $24.43\pm0.05$ & $24.09\pm0.04$ & $23.82\pm0.03$ & $22.63\pm0.01$ & $21.83\pm0.00$ & \nodata \\
 4173 & $3\;32\;41.24$ & $-27\;43\;09.7$ & $>\!27.57$     & $26.31\pm0.30$ & $25.24\pm0.19$ & $24.48\pm0.11$ & $23.91\pm0.07$ & $23.29\pm0.03$ & $22.65\pm0.02$ & \nodata \\
 4324 & $3\;32\;31.32$ & $-27\;43\;16.1$ & $>\!27.48$     & $26.23\pm0.29$ & $24.65\pm0.12$ & $23.85\pm0.07$ & $23.23\pm0.04$ & $22.25\pm0.01$ & $21.77\pm0.01$ & \nodata \\
 4327 & $3\;31\;59.71$ & $-27\;43\;15.5$ & $>\!28.40$     & $28.48\pm1.29$ & $26.81\pm0.47$ & $26.89\pm0.61$ & $25.90\pm0.36$ & $25.47\pm0.17$ & $24.37\pm0.08$ & \nodata \\
 4534 & $3\;32\;29.99$ & $-27\;43\;22.7$ & $>\!28.07$     & $27.44\pm0.51$ & $25.42\pm0.13$ & $24.37\pm0.06$ & $23.84\pm0.04$ & $22.74\pm0.01$ & $22.26\pm0.01$ & \nodata \\
 4648 & $3\;32\;26.10$ & $-27\;43\;26.7$ & $28.39\pm1.28$ & $26.60\pm0.23$ & $25.47\pm0.13$ & $24.77\pm0.08$ & $24.29\pm0.07$ & $23.49\pm0.02$ & $22.92\pm0.02$ & \nodata \\
 4846 & $3\;32\;25.93$ & $-27\;43\;31.1$ & $>\!28.28$     & $>\!28.37$     & $26.45\pm0.31$ & $26.23\pm0.31$ & $25.28\pm0.17$ & $24.08\pm0.04$ & $23.67\pm0.03$ & \nodata \\
 4921 & $3\;32\;02.44$ & $-27\;43\;35.8$ & $>\!28.03$     & $>\!28.14$     & $>\!27.58$     & $26.98\pm0.75$ & $27.88\pm1.69$ & $26.13\pm0.23$ & $24.91\pm0.10$ & \nodata \\
 5410 & $3\;32\;32.11$ & $-27\;43\;55.3$ & $>\!27.22$     & $26.33\pm0.38$ & $25.26\pm0.24$ & $25.83\pm0.50$ & $25.53\pm0.43$ & $23.87\pm0.06$ & $23.45\pm0.05$ & \nodata \\
 5529 & $3\;32\;01.77$ & $-27\;44\;01.1$ & $27.56\pm0.88$ & $28.02\pm1.16$ & $25.76\pm0.24$ & $24.76\pm0.12$ & $24.39\pm0.09$ & $23.46\pm0.03$ & $22.87\pm0.02$ & \nodata \\
 5685 & $3\;32\;12.78$ & $-27\;44\;07.7$ & $27.98\pm1.78$ & $27.70\pm1.10$ & $25.17\pm0.18$ & $24.95\pm0.18$ & $24.17\pm0.09$ & $23.44\pm0.03$ & $23.06\pm0.03$ & \nodata \\
 5735 & $3\;32\;02.83$ & $-27\;44\;09.7$ & $>\!28.05$     & $26.80\pm0.29$ & $26.88\pm0.52$ & $26.52\pm0.45$ & $26.10\pm0.33$ & $25.80\pm0.17$ & $24.66\pm0.08$ & \nodata \\
 6845 & $3\;32\;06.57$ & $-27\;45\;14.0$ & $>\!27.57$     & $27.81\pm1.15$ & $25.93\pm0.34$ & $25.06\pm0.18$ & $24.80\pm0.15$ & $23.44\pm0.03$ & $22.69\pm0.02$ & \nodata \\
 7202 & $3\;32\;06.40$ & $-27\;45\;54.7$ & $27.10\pm0.61$ & $25.76\pm0.15$ & $24.78\pm0.10$ & $23.67\pm0.05$ & $23.16\pm0.03$ & $22.37\pm0.01$ & $21.71\pm0.01$ & X-ray ID 82$^\ddagger$ \\
\hline
\multicolumn{11}{l}{$^\dagger$Coordinates refer to the J2000 epoch.}\\
\multicolumn{11}{l}{$^\ddagger$X-ray identifications are from \citet{luo08}.}
\end{tabular*}
\end{sidewaystable*}

\begin{figure}
\epsscale{1.20}
\plotone{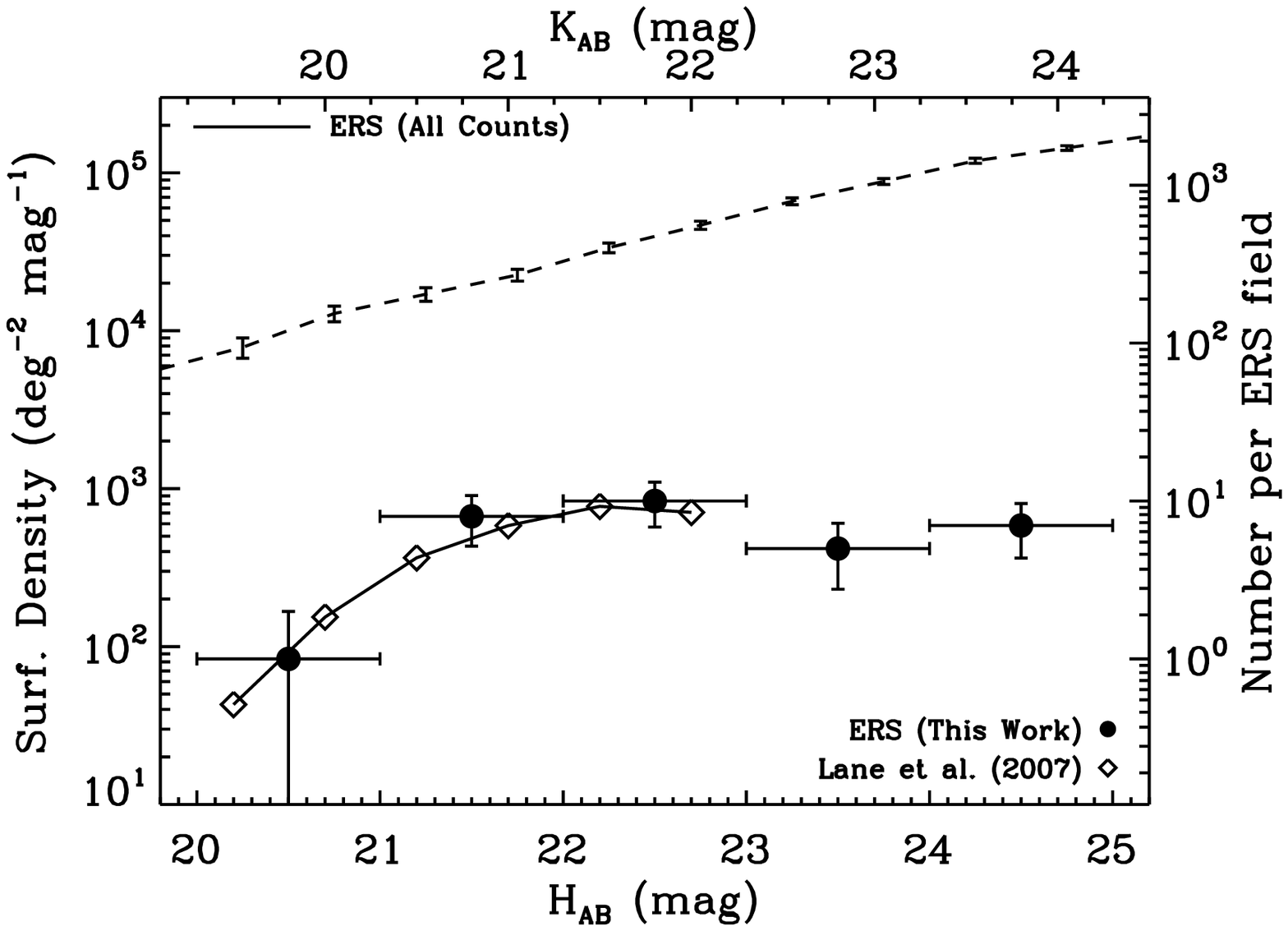}
\caption{$BzH$  galaxy  counts.   We  show  the  differential  surface
  density of galaxies selected based on the usual $BzK$ color criteria
  \citep{daddi04}       and       a       typical       color       of
  $\left<(H-K)\right>\!=\!0.7$~mag, assuming an instantaneous burst of
  star  formation and  passive evolution  from  $z_{\rm form}\!=\!10$.
  For comparison, we show  the $BzK$ object counts from \citet{lane07}
  from the UKIDSS/UDS survey as  open symbols, which were derived from
  a strict $BzK$ selection.  The upper axis is simply a transformation
  using  the assumed  $\left<(H-K)\right>$  color and  the right  axis
  applies to our objects.  To demonstrate that the apparent plateau in
  the  $BzH$ counts $H_{\rm  AB}\!\sim\!22$~mag is  not due  to simple
  survey completeness,  we show the  total source counts from  the ERS
  data as  a dashed  line \citep{win10}.  The  uncertainties presented
  here reflect the Poisson variation  in the object counts, and do not
  include any contribution from cosmic variance.\label{counts}}
\end{figure}

We  show the  source counts  of $BzH$  galaxies in  \fig{counts}.  The
counts  plateau around $H\!\simeq\!22.5$~mag,  which is  roughly 4~mag
{\it brighter}  than the formal ERS  completeness limit \citep{win10},
suggesting  that  the  faint-end   of  their  luminosity  function  is
relatively     flat    or     declining.      At    the     bright-end
($H\!\lesssim\!22$~mag),  our  survey  and modified  p$BzK$  selection
method produces source counts consistent with \citet{lane07}, based on
the UKIRT Infrared  Deep Sky Survey (UKIDSS) Ultra  Deep Survey (UDS).
While the  UKIDSS/UDS survey is much  wider ($\sim\!0.6$~deg$^2$), the
ERS  data  pushes  $\gtrsim\!2$~AB~mag   deeper,  into  a  regime  not
routinely possible from the ground.

\section{Analysis} \label{analysis}
\subsection{Surface Brightness Models} \label{surf}

To determine the  rest frame optical morphologies of  our galaxies, we
model  the two dimensional  light distribution  in the  $H$-band using
\galfit\ \citep{peng}. We  determine an empirical point-spread function
(PSF) from  a median  stack of \nstar~stars  identified in  the field,
based  on  their  full-width  at half-maximum  (FWHM)  and  brightness
determined   by   \sex\    \citep{win10}.    We   fit   the   standard
\sersic\ profile:
\begin{equation} \label{sersprof}
\Sigma(r)=\Sigma_e\,e^{-b_n\left[(r/r_e)^n-1\right]},
\end{equation}
where $r_e$ is the effective radius\footnote{We refer to the effective
  radius as $r_e$ when in angular units and as $R_e$ when converted to
  physical  units.},  $\Sigma_e$  is  the surface  brightness  at  the
effective radius, $n$  is the \sersic\ index, and  $b_n$ is a constant
found by  numerically solving $\Gamma(2n)\!=\!2\gamma(2n,b_n)$.  Since
\galfit\ is minimizing a  goodness-of-fit, modeling the uncertainty in
each pixel is  critical for meaningful estimate of  the uncertainty in
each of the  fit parameters.  Therefore, we transform  the weight maps
produced by \multidrizzle\ to  uncertainty maps and explicitly include
the shot noise  of the objects.  We perform all  fits with \galfit\ in
units of counts  using the $H$-band mosaic effective  exposure time of
5017.7~seconds.

We excise a  $81~\mbox{pix}\times81~\mbox{pix}$ stamp centered on each
galaxy; this  size was  chosen as a  compromise between  a sufficient
number of sky  pixels for a robust sky estimation  by \galfit, and the
number  of  (generally  unassociated)  nearby  galaxies.   Neighboring
galaxies  can bias  the parameter  estimation of  the  primary galaxy,
therefore  one  must carefully  mask  out  the  unmodeled objects,  or
simultaneously  fit all  the objects  in the  stamp.  We  opt  for the
latter, since it avoids the  ambiguities in pixel masking, and permits
the flux  in a  given pixel to  be represented  as the sum  of several
independent components.  However  without good initial conditions, the
\galfit\ algorithm  may not converge  to a meaningful  solution, since
the  number of degrees-of-freedom  can be  very large.   Therefore, to
estimate better  initial conditions for  each galaxy in the  stamp and
ultimately ensure  convergence in the final  simultaneous solution, we
fit the two-dimensional light profiles  of the primary $BzH$ galaxy in
question  and any  neighboring galaxies  in a  multi-stage  process as
follows:
\begin{enumerate}
\item We identify all neighboring galaxies and their associated pixels
  with  \sex, using  the same  settings mentioned  in \S~\ref{sexcat}.
  Any  sources with  $\texttt{FWHM\_IMAGE}_H\!\leq\!1$  are eliminated
  from the \sex\ segmentation maps and object catalogs, since they are
  likely not galaxies.
\item The pixels of any galaxy whose isophotes are truncated (based on
  the  \sex\ flags)  are masked,  by setting  the uncertainty  maps to
  $10^{10}$~ADU.   These  galaxies are  no  longer  considered in  the
  \galfit\ process.
\item  We  model  the  light  distribution of  each  remaining  galaxy
  (including the primary $BzH$ galaxy) individually, while masking all
  the pixels associated with  every other galaxy.  For any neighboring
  galaxy  with semi-minor  axis  of $\texttt{B\_IMAGE}_H\!\leq\!1$~pix
  from \sex, we switch from fitting a \sersic\ profile to a PSF model,
  in order  to eliminate degenerate  degrees of freedom.   Our results
  are robust to the choice  of semi-minor axis limit, provided that we
  do not  permit it to  be larger than  the size of the  empirical PSF
  (discussed in more detail below).  It is important to note that this
  step is  only present to get reasonably  accurate initial conditions
  for subsequent simultaneous object fitting.
\item  We refit  a \galfit\  model,  which contains  a combination  of
  point sources and \sersic\  profiles, to the stamp as  a whole using
  the results from the previous step as the initial guesses.
\end{enumerate}
In a few cases, we manually masked diffraction spikes or stellar halos
clearly associated with foreground  stars, which were just outside the
field-of-view  of  the  stamp  before  proceeding  through  the  above
procedure.

For  the individual and  simultaneous fits,  we placed  constraints on
various \galfit\  parameters to the  algorithm from diverging  into an
unphysical regime. We constrained the  centroid of any component to be
within the $\pm\!2\sigma$  of the centroid determined by  \sex, and the
total         magnitude         to        be         $35\!\leq\!H_{\rm
  tot}\!\leq\!\texttt{MAG\_AUTO}_H-2$~mag.     In    general,    these
constraints  are so  weak  that they  generally  play no  role in  the
fitting  whatsoever, but  they  reduce the  sensitivity  to the  pixel
masking  with the  segmentation  maps. We  additionally constrain  the
\sersic\ index and effective radius to be $0.01\!\leq\!n\!\leq\!8$ and
$0.01\!\leq\!r_e/\texttt{A\_IMAGE}\!\leq\!5$,   respectively.    These
constraints are considerably stronger, and we recognize that the model
is likely incorrect when \galfit\  converges to a solution which is on
these boundaries.   These cases  are rare, and  generally a  sign that
additional   astrophysical  components  are   needed  (e.g.~bulge/disk
separately,  nuclear point  sources, or  merger signatures),  that the
frame  was  inappropriately sized,  that  the  object was  unresolved,
and/or that there was some  additional light component or image defect
present in  the image (e.g.~diffraction spikes,  stellar halos, and/or
cosmic rays).

As our primary interest here is  on the sizes of these galaxies, it is
imperative that  we ensure the effective radii  are robustly measured.
While  \galfit\  reliably  determines  the  random  uncertainty  which
follows from  the maximum  likelihood analysis, the  total uncertainty
should  include   a  systematic  term   as  well.   To   estimate  the
contribution from  the systematic uncertainty, we construct  a grid of
simulated  galaxies   with  brightnesses  $20\!\leq\!H\!\leq\!25$~mag,
effective   radii   $0.5\!\leq\!r_e\!\leq\!5.5$~pix,   and   a   fixed
\sersic\ index  of $n\!=\!4$.  These  galaxies are convolved  with the
PSF and  embedded in a blank  region of the $H$-band  mosaic.  We then
fit these simulated galaxies  with \galfit\ using the above procedure,
and  find  that  the   effective  radii  are  generally  uncertain  by
$\sim\!10$\%,   which   is    somewhat   brightness   dependent.    We
quadratically   add  this   systematic  uncertainty   to   the  random
uncertainty determined by \galfit.

\begin{figure}
\epsscale{1.20}
\plotone{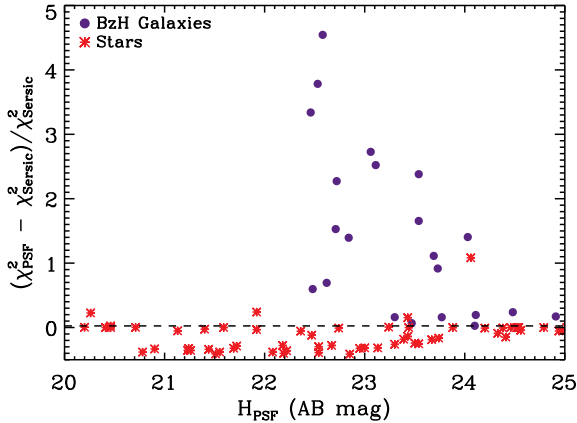}
\caption{Comparison  of   PSF  and  \sersic\  models.    We  show  the
  fractional difference between  the $\chi^2$ goodness-of-fit measures
  from \galfit\ for  the PSF and \sersic\ models.   The $BzH$ galaxies
  are shown as  filled blue points, while the  red asterisks represent
  \nstar~ERS  point sources, presumed  to be  stars.  For  \galfit\ to
  optimally fit  an unresolved  source with a  PSF, it will  drive the
  effective radius  and \sersic\ index to unphysical  regimes, {\it de
    facto}  fitting  a  PSF.   Therefore,  the  goodness-of-fit  of  a
  \sersic\ model and  a PSF should be roughly  equal.  However, if one
  were  to fit a  PSF to  a resolved  object, then  there should  be a
  noticeable  increase in  the goodness-of-fit  statistic.  We  adopt a
  critical  value  of   $F_{\rm  crit}\!=\!0.01$  \citep{bond09},  see
  \S~\ref{surf} for  more details.  Of  the \nstar~stars used  in this
  test, only  4~have $F\!>\!F_{\rm crit}$ suggesting  this is reliable
  way to classify unresolved objects. \label{ftest}}
\end{figure}

Many of the these red galaxies are very small and, even with the space
based imaging, may  still be unresolved.  Therefore it  is critical to
properly  identify  which  galaxies  are resolved  and  have  reliable
effective radii measurements.  We begin by swapping the fully variable
\sersic\ model for the primary galaxy with a pure PSF model, which can
only  vary   in  position  and   brightness.   However,  as   we  have
simultaneously fit  every object in our postage  stamps, we anticipate
that  \galfit\ may  incorrectly change  the parameters  of neighboring
(unrelated)  galaxies  to  compensate  for  the  poor  primary  model,
particularly  in   the  case  of  a  well   resolved  primary  galaxy.
Therefore, when we  use the PSF model for the  primary galaxy, we hold
the parameters  of the neighboring  sources fixed at the  values found
previously by \galfit.  We  now compare the goodness-of-fit statistics
for these \sersic\ and PSF models by considering the quantity:
\begin{equation} \label{fdef}
F=\frac{\chi^2_{\mbox{PSF}}-\chi^2_{\mbox{\sersic}}}{\chi^2_{\mbox{\sersic}}}, 
\end{equation}
and  expect that  sources  with low  values  of $F$  are equally  well
characterized by a PSF model  as by the more complex \sersic\ profile.
We calibrate  this quantity by  computing the $F$-values of  the known
ERS stars, which were used to  derive the empirical PSF used above.  In
\fig{ftest},   we  show   the  $F$-values   as  a   function   of  the
\galfit-derived   PSF  magnitude,   with  the   galaxies   plotted  as
filled-blue points  and the stars as  red asterisks.  The  trend is as
expected:   nearly    all   Galactic    stars   can   be    found   at
$-0.5\!\lesssim\!F\!\lesssim\!0$,   while  the   $BzH$   galaxies  are
generally at $F\!\gtrsim\!0$  which depends on brightness.  Therefore,
we define objects which can be  equally characterized by a PSF as by a
\sersic\  fit as  having $F\!\leq\!F_{\rm  crit}$, while  objects with
$F\!>\!F_{\rm crit}$ are  more extended than the known  ERS stars.  We
adopt $F_{\rm crit}\!=\!0.025$ and  note that only 4/\nstar~stars have
$F\!>\!F_{\rm  crit}$, which  is consistent  with  \citet{bond09}.  We
give the \galfit\ results in \tab{bzh_galfit}.

\begin{table*}
\caption{$BzH$ \galfit\ Results}
\label{bzh_galfit}
\begin{tabular*}{0.98\textwidth}
  {@{\extracolsep{\fill}}lccccl}
\hline\hline
\multicolumn{1}{c}{ID} & \multicolumn{1}{c}{$r_e$} & \multicolumn{1}{c}{$n^\dagger$} & \multicolumn{1}{c}{$\chi^2_{\nu}$} & \multicolumn{1}{c}{$F^\ddagger$} & \multicolumn{1}{c}{Notes}\\
\multicolumn{1}{c}{$ $} & \multicolumn{1}{c}{(arcsec)} & \multicolumn{1}{c}{$ $} & \multicolumn{1}{c}{$ $} & \multicolumn{1}{c}{$ $} & \multicolumn{1}{c}{$ $}\\
$ $ & $ $ & $ $ & $ $ & $ $ & $ $ \\
\hline
  408 & $0.04\pm0.00$ & $4.72\pm0.45$ & $0.487$ & $ 0.160$ & \nodata \\
  606 & $0.09\pm0.01$ & $6.92\pm0.92$ & $0.404$ & $ 0.158$ & \nodata \\
 1696 & $0.41\pm0.02$ & $1.21\pm0.10$ & $0.356$ & $ 0.170$ & \nodata \\
 2227 & $0.35\pm0.01$ & $2.93\pm0.09$ & $0.364$ & $ 2.517$ & \nodata \\
 2377 & $0.16\pm0.01$ & $4.17\pm1.01$ & $0.322$ & $ 0.049$ & \nodata \\
 2749 & $0.51\pm0.02$ & $0.57\pm0.05$ & $1.849$ & $ 0.191$ & \nodata \\
 2750 & $1.03\pm0.19$ & $5.65\pm0.58$ & $1.891$ & $ 0.591$ & \nodata \\
 2871 & $0.03\pm0.00$ & $8.00\pm2.76$ & $0.567$ & $ 0.023$ & unresolved \\
 3000 & $0.60\pm0.01$ & $8.00\pm0.14$ & $0.814$ & $ 9.440$ & tidal tail \\
 3152 & $0.14\pm0.00$ & $4.66\pm0.12$ & $0.515$ & $ 3.323$ & \nodata \\
 3237 & $0.22\pm0.01$ & $7.54\pm0.30$ & $0.432$ & $ 2.268$ & \nodata \\
 3360 & $0.11\pm0.00$ & $7.58\pm0.46$ & $0.466$ & $ 1.382$ & \nodata \\
 3376 & $0.15\pm0.01$ & $2.21\pm0.44$ & $0.349$ & $ 0.054$ & \nodata \\
 3471 & $0.08\pm0.00$ & $4.40\pm0.16$ & $0.478$ & $ 1.525$ & \nodata \\
 3551 & $0.44\pm0.01$ & $2.74\pm0.11$ & $0.614$ & $ 0.851$ & additional nuclear point source was fit \\
 3812 & $0.16\pm0.00$ & $3.12\pm0.07$ & $0.486$ & $ 3.767$ & \nodata \\
 4148 & $0.04\pm0.00$ & $5.73\pm0.32$ & $0.443$ & $ 0.690$ & \nodata \\
 4173 & $0.37\pm0.00$ & $0.98\pm0.02$ & $0.359$ & $ 2.373$ & \nodata \\
 4324 & $0.25\pm0.00$ & $3.27\pm0.08$ & $0.404$ & $ 4.536$ & \nodata \\
 4327 & $0.03\pm0.02$ & $2.96\pm3.62$ & $0.411$ & $ 0.000$ & unresolved \\
 4534 & $0.14\pm0.00$ & $2.21\pm0.07$ & $0.388$ & $ 2.718$ & \nodata \\
 4648 & $0.17\pm0.00$ & $2.08\pm0.10$ & $0.355$ & $ 0.913$ & \nodata \\
 4846 & $0.16\pm0.00$ & $1.30\pm0.13$ & $0.506$ & $ 0.232$ & \nodata \\
 4921 & $0.19\pm0.08$ & $8.00\pm4.50$ & $0.337$ & $ 0.012$ & unresolved \\
 5410 & $0.00\pm1.00$ & $2.64\pm1.00$ & $0.364$ & $-0.009$ & likely three distinct clumps \\
 5529 & $0.29\pm0.00$ & $2.63\pm0.12$ & $0.369$ & $ 1.107$ & \nodata \\
 5685 & $0.48\pm0.00$ & $0.36\pm0.02$ & $0.386$ & $ 1.399$ & \nodata \\
 5735 & $0.08\pm0.01$ & $8.00\pm4.15$ & $0.349$ & $ 0.012$ & unresolved \\
 6845 & $0.42\pm0.01$ & $1.33\pm0.04$ & $0.385$ & $ 1.648$ & \nodata \\
 7202 & $0.30\pm0.00$ & $2.29\pm0.04$ & $0.434$ & $ 5.825$ & \nodata \\
$ $ & $ $ & $ $ & $ $ & $ $ & $ $ \\
\hline
\multicolumn{6}{l}{$^\dagger$The \sersic\ index in \eqn{sersprof}.}\\
\multicolumn{6}{l}{$^\ddagger$The fractional difference between the goodness-of-fit for the PSF and \sersic\ models (see \S~\ref{surf} for more details).}
\end{tabular*}
\end{table*}

\subsection{Additional Photometry} \label{addphot}

Our $BzH$ criteria are  designed to select passively evolving galaxies
at  $z\!\sim\!1.5$,  consequently  the  \hst\ photometry  covers  only
$\lambda_{\rm   rest}\!\lesssim\!6000$~\AA.    We   can   extend   our
wavelength  coverage to  $\lambda_{\rm  rest}\!\sim\!3~\mu$m with  the
$K_s$-band   imaging    from   the   {\it    Very   Large   Telescope}
\citep[\vlt;][]{retz} and the four IRAC channels from the {\it Spitzer
  Space  Telescope}  (\sst; PI:  M.~Dickinson).   However the  notably
lower spatial  resolution ($0\farcs7-2''$)  of these images  demands a
different  approach for measuring  the flux,  as source  confusion can
significantly bias the photometry.

By  inspection  of the  images,  it is  clear  that  our galaxies  are
unresolved in  the \vlt\ and  \sst\ data.  Therefore, we  obtain total
magnitudes  from  a  \texttt{GalFit}  model with  a  similar  approach
discussed  in \S~\ref{surf},  with  a few  simplifications. First,  we
assume that all  the sources in a given  postage stamp are unresolved,
and are therefore  ideally modeled by a PSF.  Second,  we do not allow
the centroids  of all objects to  vary more than  $\pm\!1$~pix (of the
\vlt\ or  \sst\ images).  Third, we only  perform simultaneous fitting
with initial magnitudes given by the $H$-band measurements, as in many
cases  multiple \hst\  sources  are  blended into  a  single \vlt\  or
\sst\ source.

\subsection{Stellar Populations and Photometric Redshifts} \label{photz}

We fit the 15-band (\hst, \vlt, and \sst) photometry with a library of
stellar  population synthesis models  to simultaneously  determine the
stellar mass, population age, and redshift with our own software.  Our
model grid consists  of a four dimensional parameter  space spanned by
redshift  ($z$), stellar population  age\footnote{We impose  the usual
  self-consistency constraint  that the a  galaxy be younger  than the
  age of the Universe.}   ($t$), $V$-band extinction ($A_V$), and star
formation  timescale  ($\tau$)  for  an exponentially  declining  star
formation history.   We assume  solar metallicity, a  Salpeter initial
mass function,  and adopt the  CB07 population synthesis  models.  The
allowed  photometric  redshifts  ranged from  $0\!\leq\!z\!<\!7$  with
$\Delta z\!=\!0.01$, the ages  adopted by \citet{bolz}, extinctions of
$0\!\leq\!A_V\!\leq\!2$~mag  with $\Delta  A_V\!=\!0.2$~mag,  and star
formation timescales of $\tau\!\in\![10^{-3},1,2,3,5,15,30,10^3]$~Gyr.
We include a systematic uncertainty on the observed fluxes of the form
$\sigma_f\!=\!\alpha\times f$,  where we adopt  $\alpha\!=\!0.1$, 0.1,
0.2, and 0.2 for the ACS,  WFC3, \vlt, and \sst\ data, respectively to
account   for  uncertainties   in  the   zeropoints,   templates,  and
measurement  approach (whether  \sex\  or \galfit,  for  the \hst\  or
\vlt/\sst\  data).   We compute  the  $1\sigma$  uncertainties on  the
stellar  population parameters  (e.g.~mass, age,  etc.)  by  use  of a
simple Monte Carlo calculation.  For  each band for a given galaxy, we
draw a normal  random variable with mean and  standard deviation equal
to the  flux and flux  uncertainty (without the the  systematic term).
By repeating for  many iterations, we build up  a distribution of each
stellar population parameter and  take the mean and standard deviation
as the measured quantity and $1\sigma$ uncertainties, respectively.

\begin{figure}
\epsscale{1.20}
\plotone{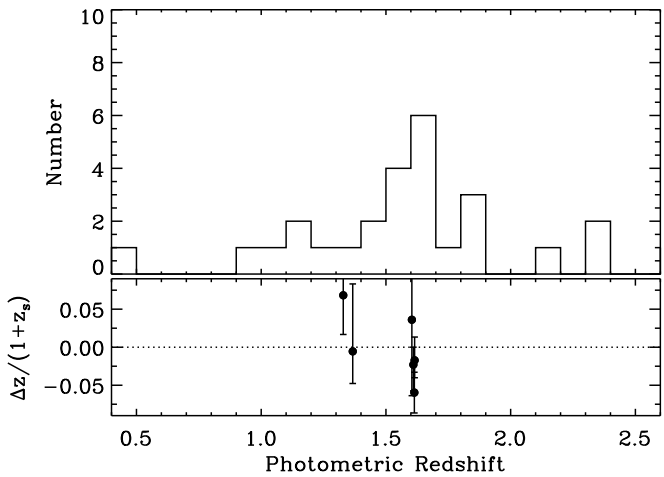}
\caption{Photometric  redshift distribution.  In  the upper  panel, we
  show the photometric redshift  distribution derived by the procedure
  described  in  \S~\ref{photz}.  As  expected,  our  sample of  $BzH$
  galaxies  is generally  located at  $z_{\rm phot}\!=\!1.6$.   In the
  lower panel, we show  the fractional difference between our redshift
  estimates and spectroscopic redshifts, where available. Based on the
  RMS  on the  fractional  differences, we  estimate  that $(1+z)$  is
  accurate to $\sim\!4.6$\%.\label{zdist}}
\end{figure}

We  apply  this  approach  to  the complete  sample  of  $\nbzh$~$BzH$
galaxies.    In  \fig{zdist},   we  show   the   photometric  redshift
distribution  and the  comparison  to a  spectroscopic redshift,  when
available.  As expected, the median  redshift of the sample is $z_{\rm
  phot}\!=\!1.6\pm\!0.6$, where the  uncertainty reflects the standard
deviation  of  the distribution.   To  estimate  our typical  redshift
uncertainty, we  compare to  the published spectroscopic  redshifts in
the  lower  panel of  \fig{zdist}.   Based  on  the root-mean  squared
scatter of  the seven objects  with known spectroscopic  redshifts, we
estimate  our  uncertainties  in  $(1+z)$ are  $\sim\!4.6$\%.   We  do
recover the known galaxy  cluster at $z\!\sim\!1.6$ \citep{gmass}.  We
present  our resulting  photometric redshifts  and  stellar population
parameters in \tab{photztab}.

\begin{table}
\caption{Photometric Redshifts and Stellar Population Parameters}
\label{photztab}
\begin{tabular*}{0.48\textwidth}
  {@{\extracolsep{\fill}}lccccc}
\hline\hline

\multicolumn{1}{c}{ID} & \multicolumn{1}{c}{$z_{\rm phot}$} & \multicolumn{1}{c}{$z_{\rm spec}$} & \multicolumn{1}{c}{Age$^\dagger$} & \multicolumn{1}{c}{$M_*^\ddagger$} & \multicolumn{1}{c}{$\chi^2_{\nu}$} \\
\multicolumn{1}{c}{$ $} & \multicolumn{1}{c}{$ $} & \multicolumn{1}{c}{$ $} & \multicolumn{1}{c}{(Gyr)} & \multicolumn{1}{c}{($10^{11}~M_{\odot}$)} & \multicolumn{1}{c}{$ $}\\
$ $&$ $&$ $&$ $&$ $&$ $\\
\hline
  408 & $1.59_{-0.04}^{+0.05}$ & \nodata & $0.5_{-0.0}^{+0.0}$ & $ 0.4\pm 0.0$ &  1.5 \\
  606 & $1.87_{-0.20}^{+0.24}$ & \nodata & $0.7_{-0.2}^{+0.3}$ & $ 0.4\pm 0.0$ &  3.2 \\
 1696 & $4.12_{-3.58}^{+0.56}$ & \nodata & $0.4_{-0.2}^{+3.1}$ & $11.3\pm 1.1$ &  2.4 \\
 2227 & $1.27_{-0.13}^{+0.11}$ & \nodata & $4.5_{-1.0}^{+1.0}$ & $ 1.3\pm 0.1$ &  5.0 \\
 2377 & $2.36_{-0.54}^{+0.13}$ & \nodata & $0.5_{-0.4}^{+3.0}$ & $ 0.1\pm 0.0$ &  0.6 \\
 2749 & $1.85_{-0.09}^{+0.27}$ & \nodata & $0.1_{-0.0}^{+0.0}$ & $ 0.8\pm 0.0$ &  6.0 \\
 2750 & $1.87_{-0.09}^{+0.11}$ & \nodata & $0.5_{-0.4}^{+0.2}$ & $ 0.8\pm 0.0$ &  8.4 \\
 2871 & $1.54_{-0.09}^{+0.08}$ & \nodata & $0.7_{-0.4}^{+0.0}$ & $ 0.2\pm 0.0$ &  1.9 \\
 3000 & $1.15_{-0.04}^{+0.42}$ & \nodata & $5.5_{-5.0}^{+0.0}$ & $ 5.1\pm 0.2$ &  4.1 \\
 3152 & $1.38_{-0.10}^{+0.21}$ &  1.367  & $4.5_{-3.5}^{+0.0}$ & $ 2.7\pm 0.1$ &  0.3 \\
 3237 & $1.66_{-0.06}^{+0.08}$ &  1.615  & $0.7_{-0.2}^{+0.0}$ & $ 1.1\pm 0.1$ &  4.0 \\
 3360 & $1.63_{-0.09}^{+0.44}$ & \nodata & $1.0_{-0.5}^{+1.0}$ & $ 1.2\pm 0.1$ &  1.4 \\
 3376 & $1.49_{-0.11}^{+0.09}$ & \nodata & $0.3_{-0.1}^{+0.5}$ & $ 0.1\pm 0.0$ &  4.3 \\
 3471 & $1.67_{-0.09}^{+0.06}$ &  1.610  & $1.0_{-0.3}^{+0.0}$ & $ 0.7\pm 0.0$ &  2.5 \\
 3551 & $1.60_{-0.10}^{+0.06}$ & \nodata & $1.0_{-0.3}^{+0.4}$ & $ 2.2\pm 0.1$ &  0.3 \\
 3812 & $1.77_{-0.07}^{+0.07}$ &  1.614  & $1.0_{-0.0}^{+0.0}$ & $ 1.0\pm 0.0$ &  4.9 \\
 4148 & $2.38_{-0.05}^{+0.07}$ & \nodata & $0.4_{-0.0}^{+0.0}$ & $ 1.3\pm 0.1$ &  2.8 \\
 4173 & $0.98_{-0.67}^{+0.31}$ & \nodata & $3.5_{-2.1}^{+6.0}$ & $ 0.3\pm 0.0$ &  4.8 \\
 4324 & $1.55_{-0.09}^{+0.10}$ & \nodata & $1.0_{-0.3}^{+0.4}$ & $ 0.9\pm 0.0$ &  2.5 \\
 4327 & $3.15_{-0.77}^{+1.10}$ & \nodata & $0.2_{-0.1}^{+2.1}$ & $ 1.9\pm 0.2$ &  2.3 \\
 4534 & $1.51_{-0.26}^{+0.17}$ &  1.604  & $1.4_{-0.4}^{+3.1}$ & $ 0.9\pm 0.0$ &  2.8 \\
 4648 & $0.41_{-0.10}^{+0.61}$ & \nodata & $8.5_{-7.1}^{+1.0}$ & $ 0.0\pm 0.0$ &  2.8 \\
 4846 & $1.63_{-0.30}^{+0.10}$ & \nodata & $0.5_{-0.3}^{+4.0}$ & $ 0.3\pm 0.0$ &  7.1 \\
 4921 & $3.37_{-1.69}^{+1.38}$ & \nodata & $0.7_{-0.5}^{+2.8}$ & $ 2.3\pm 0.3$ &  2.0 \\
 5410 & $2.12_{-0.37}^{+0.17}$ & \nodata & $0.4_{-0.3}^{+3.1}$ & $ 0.2\pm 0.0$ &  7.9 \\
 5529 & $1.03_{-0.35}^{+0.24}$ & \nodata & $4.5_{-3.5}^{+3.0}$ & $ 0.4\pm 0.0$ &  3.5 \\
 5685 & $1.41_{-0.14}^{+0.13}$ & \nodata & $1.4_{-0.4}^{+1.2}$ & $ 0.2\pm 0.0$ &  9.3 \\
 5735 & $2.64_{-0.15}^{+0.31}$ & \nodata & $0.3_{-0.2}^{+2.0}$ & $ 0.1\pm 0.0$ &  3.0 \\
 6845 & $1.67_{-0.37}^{+0.64}$ & \nodata & $2.6_{-1.9}^{+1.9}$ & $ 0.9\pm 0.1$ &  5.2 \\
 7202 & $1.17_{-0.12}^{+0.11}$ &  1.329  & $2.6_{-0.3}^{+2.9}$ & $ 1.7\pm 0.1$ &  2.0 \\
$ $&$ $&$ $&$ $&$ $&$ $\\
\hline
\multicolumn{6}{l}{$^\dagger$Population age assuming an exponetial star formation history.}\\
\multicolumn{6}{l}{$^\ddagger$Stellar mass.}
\end{tabular*}
\end{table}

\section{$BzH$ Galaxy Size Evolution}\label{size}

\begin{figure}
\epsscale{1.20}
\plotone{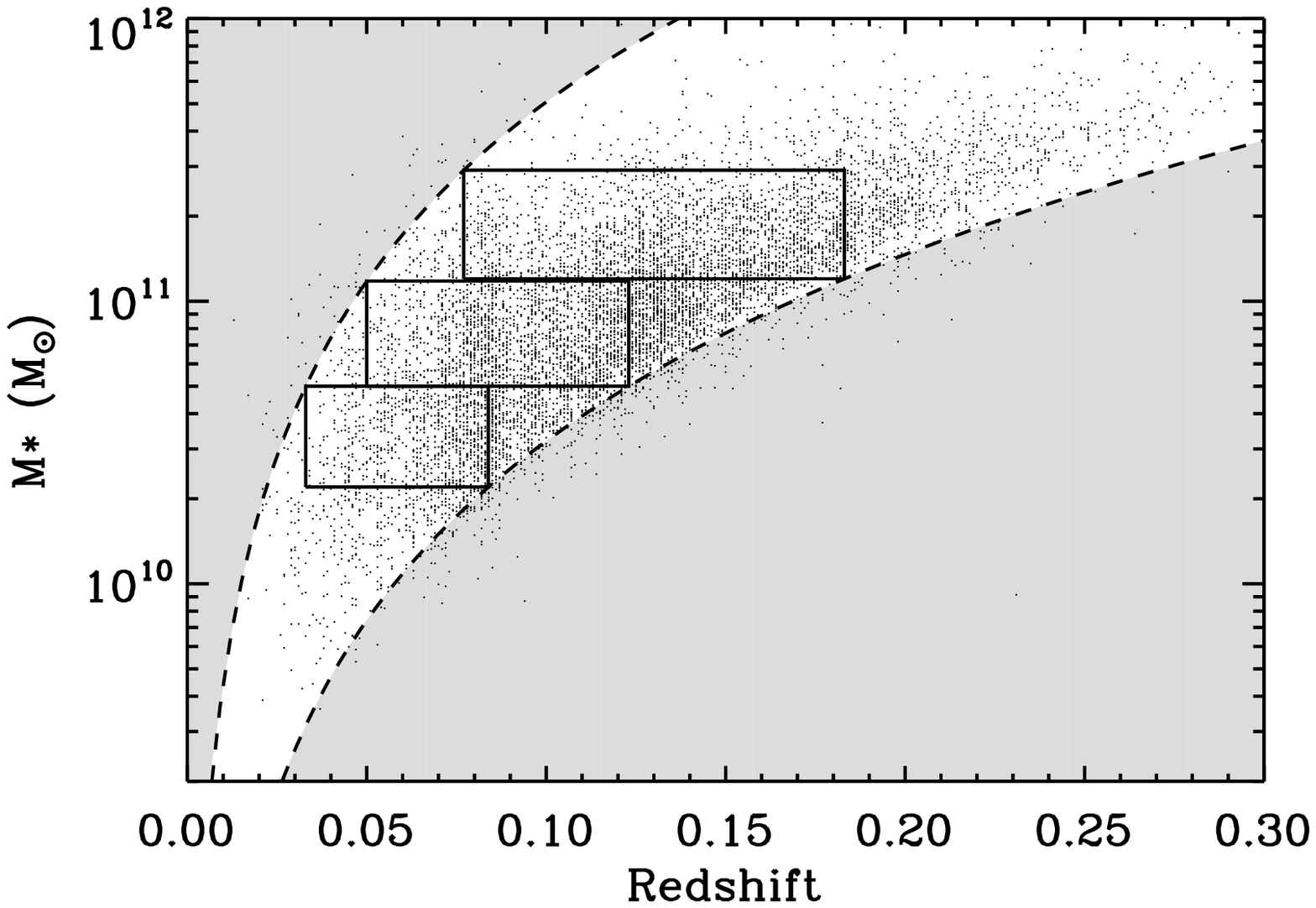}
\caption{Selection of SDSS galaxy sample.  We show the stellar mass as
  a  function  of spectroscopic  redshift  for  the early-type  galaxy
  sample of \citet{bern03} drawn  from the SDSS-EDR.  The stellar mass
  estimates  were derived  from stellar  population fits  to  the SDSS
  photometry  \citep{salim07},  and are  similar  in  nature to  those
  described in  \S~\ref{photz}.  The dashed lines  indicate the bright
  ($i'\!\leq\!14$~mag)  and  faint  ($i'\!\geq\!17$~mag)  completeness
  limits set by detector  saturation and the \citet{bern03} selection,
  respectively.   The  boxes show  our  volume-limited selections  for
  proper low-redshift  comparisons.  If  such limits are  not imposed,
  then an artificial trend in effective radius with redshift will arise
  from a Malmquist-type bias.\label{sdss}}
\end{figure}

Based on the photometric redshift estimates and the spectroscopic data
(where available), the $BzH$ selection reliably identifies galaxies in
the interval $\left<z\right>\!\sim\!1.6\pm0.6$.   However to study the
evolution  of their  sizes with  redshift,  we must  compare to  other
similarly  selected samples.   The  high redshift  ($z\!\gtrsim\!1.5$)
samples are  generally derived  from similar color  criteria presented
here, and  have effective radii measured  in the $H$-band,  for a rest
frame wavelength  of $\lambda_{\rm rest}\!\sim\!6500$~\AA.   To ensure
fair comparisons with lower  redshift samples, we require similar rest
frame  sizes.   For  the low  redshift  data,  we  use the  sample  of
8666~early-type  galaxies  at   $z\!\sim\!0.2$  with  effective  radii
measured  in  the $i'$-band  \citep{bern03}  selected  from the  Sloan
Digital  Sky Survey,  Early Data  Release \citep[SDSS-EDR;][]{york00}.
The  stellar masses  for the  7th Data  Release\footnote{Obtained from
  http://www.mpa-garching.mpg.de/SDSS/DR7/.}  galaxies were determined
following \citet{salim07}.  By cross matching these samples, we obtain
8595~galaxies for  our low-redshift comparison.  We  select three mass
ranges  which are  volume-limited based  on the  upper and  lower flux
limits imposed  by detector saturation in SDSS  and the \citet{bern03}
brightness  criterion.   In \fig{sdss}  we  show these  volume-limited
selections (black boxes) and the flux limits (dashed lines).  If these
limits  are   not  strictly  imposed,  then   an  artificial  redshift
dependence on the effective radii  will be introduced as the radii are
tightly correlated  with the stellar masses, which  roughly scale with
luminosity.  For  example, without these volume  limits, the effective
radii of the  SDSS galaxies will seem to  {\it increase} with redshift
since the survey is not sensitive to the lower mass (smaller) galaxies
at  the higher redshifts.   Eliminating this  potential Malmquist-like
bias in the low redshift  sample ensures a fair comparison between the
high and low redshift data.

\begin{figure}
\epsscale{1.20}
\plotone{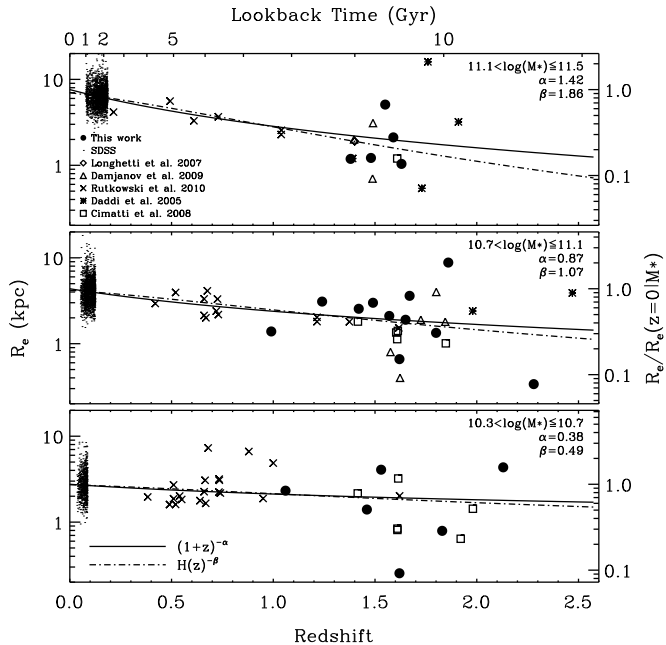}
\caption{$BzH$ galaxy  size evolution.   We show the  effective radius
  versus  redshift for the  various stellar  mass slices  described in
  \fig{sdss} and several early-type galaxy samples. We do not show the
  uncertainties for  clarity.  As solid  and dashed lines we  show the
  best     fit    model    of     $R_e\!\propto(1+z)^{-\alpha}$    and
  $R_e\!\propto\!H(z)^{-\beta}$, respectively.   While the data cannot
  rule out either model, we  show both to help illustrate the emerging
  trend:  the  increase  in  effective radii  from  $z\!\sim\!2.5$  to
  present     is      strongest     in     high      mass     galaxies
  ($M^*\!\geq\!10^{11}~M_{\odot}$).   Based on these  fits, we  give a
  tentative   estimate   for   $\alpha(M_*)$   and   $\beta(M_*)$   in
  equations~(\ref{aeqn})               and               (\ref{beqn}),
  respectively.  \label{sizefig}}
\end{figure}

To investigate the passively evolving galaxy size evolution at a fixed
stellar  mass, we  show  in  \fig{sizefig} the  effective  radii as  a
function of  redshift for this  work (large circles), the  SDSS sample
(small  dots), \citet[][diamonds]{long07}, \citet[][triangles]{dam09},
\citet[][crosses]{mjr10},        \citet[][asterisks]{dad05},       and
\citet[][squares]{cim08}.   We overplot  the two  canonical  models of
$(1+z)^{-\alpha}$ (solid line)  and $H(z)^{-\beta}$ (dot-dashed line),
where  $H(z)$  is the  Hubble  parameter,  and  present the  best  fit
parameters  in \tab{sizeevol}.   While the  data do  not significantly
favor either  model, they serve  to highlight an important  trend: the
amount  by which  galaxies  are smaller  in  the past  depends on  the
stellar   mass.   \citet{new10}  identified   a  similar   result  for
12~galaxies  with dynamical masses  determined from  measured velocity
dispersions.

Our  stellar  mass-dependent  size  evolution model  is  qualitatively
similar to  that proposed by \citet{hop09}, where  the power-law index
varies                  with                  mass                  as
$\alpha(M_*)\!\approx\!0.23\log{\left(M_*/10^9~M_{\odot}\right)}$.
However, their  model underpredicts our measured  power-law indices in
our three mass regimes, but  does give the same qualitative steepening
of $\alpha$ with mass.   Following their approach, we derive tentative
relationships for the power-law indices:
\begin{eqnarray}\label{aeqn}
\alpha(M_*)&\approx&-1.8+1.4\times\log\left(\frac{M_*}{10^9~M_{\odot}}\right)\\ \beta(M_*)&\approx&-2.3+1.8\times\log\left(\frac{M_*}{10^9~M_{\odot}}\right), \label{beqn}
\end{eqnarray}
and caution  that with  only three independent  mass bins,  these fits
should be considered preliminary at best.  However, these results give
indices consistent with the reported value of \citet{buit08}, who find
that $\alpha(M_*\!>\!10^{11}~M_{\odot})\!\approx\!1.5$.

\begin{table}
\caption{Mass-Dependent Size Evolution Models}
\label{sizeevol}
\begin{tabular*}{0.48\textwidth}
    {@{\extracolsep{\fill}}ccccc}
\hline\hline
\multicolumn{1}{c}{$\log{M_{\mathrm{low}}}$} & \multicolumn{1}{c}{$\log{M_{\mathrm{high}}}$} & \multicolumn{1}{c}{$\alpha^\dagger$} & \multicolumn{1}{c}{$\beta^\ddagger$} & \multicolumn{1}{c}{$R_e(z\!=\!0)$}\\
\multicolumn{1}{c}{($\log{M_\odot}$)} & \multicolumn{1}{c}{($\log{M_\odot}$)} & \multicolumn{1}{c}{$ $} & \multicolumn{1}{c}{$ $} & \multicolumn{1}{c}{(kpc)}\\
$ $ &$ $ &$ $ &$ $ &$ $ \\
\hline
10.3 & 10.7 &  0.38 &  0.49 & 2.76 \\
10.7 & 11.1 &  0.87 &  1.07 & 4.36 \\
11.1 & 11.5 &  1.42 &  1.86 & 7.60 \\
$ $ &$ $ &$ $ &$ $ &$ $ \\
\hline
\multicolumn{5}{l}{$^\dagger$For the $(1+z)^{-\alpha}$ model.}\\
\multicolumn{5}{l}{$^\ddagger$For the $H(z)^{-\beta}$ model.}
\end{tabular*}
\end{table}

\section{Discussion} \label{disc}

It has become relatively well established that early-type galaxies are
indeed smaller  at high redshift  than their local counterparts  for a
given stellar or  dynamical mass \citep[e.g.][]{tru09,new10}.  However
the causes  for this result are  far less clear, or  agreed upon.  The
likely  mechanisms  can  be  broadly  characterized  as  astrophysical
effects (early time major  mergers, late time minor mergers, adiabatic
expansion,  or  stellar  mass-to-light  gradients),  or  observational
biases  (underestimating  the   effective  radii,  overestimating  the
stellar masses, or incorrectly assuming that the high and low redshift
populations are directly  comparable).  Using a semi-analytical model,
\citet{hop10}  find  that the  factor  of  $\sim\!5$  increase in  the
effective  radii  over  the  last  $\sim\!10$~Gyr  for  galaxies  with
$M_*\!\geq\!10^{11}~M_{\odot}$ can be explained by a combination these
effects, with  the late-time minor  merging playing the  largest role.
However,  we  find that  lower  mass  systems  exhibit notably  weaker
redshift  evolution,  suggesting  a  different  mixture  of  the  main
processes may be  at work.  For example, the  adiabatic expansion mode
may become more critical  given the shallower gravitational potentials
in  the  low  mass  systems.

As noted  above, the {\it stellar mass-dependent  size evolution} seen
here is similar to the  findings of \citet{new10} for dynamical masses
estimated   from  velocity   dispersions,  and   has   an  interesting
consequence for the $R_e-M_*$ relation.  \citet{shen03} find that SDSS
elliptical      galaxies     follow      the      scaling     relation
$R_e\!=\!R_{e,11}(M_*/10^{11}~M_{\odot})^\gamma$,                 where
$R_{e,11}\!=\!4.16$~kpc    and    $\gamma\!=\!0.56$.    However    the
mass-dependent size evolution observed here and by \citet{new10} imply
either  a fundamental change  in the  $R_e-M$ scaling  relationship at
high  redshift (such  as  a  flattening of  the  effective radius  for
decreasing stellar mass) or a redshift dependent value of $\gamma$, in
addition to  the usual  lower value of  $R_{e,11}$.  However,  at this
stage the high redshift data cannot distinguish these two scenarios or
shed light on the cause for this change.

The red  galaxy formation paradigm  wherein mergers of  gas-rich discs
trigger intense starbursts which are later quenched by active galactic
nuclei, finally giving  way to dead spheroidal systems  spent of their
gas \citep[e.g.][and references therein]{faber07} suggests the passive
galaxies may  have come from a  population of more  active galaxies in
their  recent  past.   The  Lyman-break galaxies  (LBGs)  observed  at
$z\!\gtrsim\!3$ are a possible progenitor system.  While their stellar
masses are far less well constrained  due to the lack of rest frame IR
data, the majority of LBGs have $M_*\!\sim\!10^{10}~M_{\odot}$ between
$3\!\lesssim\!z\!\lesssim\!6$ \citep[e.g.][]{pap01,yan06}.  Therefore,
the typical LBG at  $3\!\lesssim\!z\!\lesssim\!6$ belongs in the lower
panel of \fig{sizefig}, and have effective radii of $R_e\!\sim\!1$~kpc
\citep[e.g.][]{ferg04,bou06,nph08,oesch10},  after transforming  to an
equivalent  rest frame wavelength  \citep{barden}.  The  LBGs  are then
consistent  with our $H(z)^{-\beta}$  model, possibly  suggesting that
the physical mechanism driving the  passive galaxy evolution may be at
work  for the  LBGs as  well.   Given their  increased star  formation
rates, larger gas content, and  lower total mass, the mass-loss modes,
whether   driven  by  AGN   \citep[e.g.][]{fan08}  or   stellar  winds
\citep[e.g.][]{dam09},  are   likely  more  important.    However,  we
recognize  that these samples  (LBGs and  passive galaxies)  are quite
different  and  that  the  next  generation  of  space-based  infrared
instruments (Near-Infrared Camera  and Mid-Infrared Instrument on {\it
  James Webb Space Telescope}) will provide a much clearer picture for
the high redshift ($z\!\gtrsim\!3$) size comparisons.

\section{Summary}\label{summary}

We identified \nbzh\  passively evolving galaxies to $H\!\leq\!25$~mag
from  a set of  color criteria  similar to  the p$BzK$  selection.  We
measure  rest  frame  optical  ($\lambda_{\rm  rest}\!\sim\!6500$~\AA)
effective  radii as two-dimensional  fits to  the $H$-band  image.  By
comparing with several other  comparable samples at various redshifts,
we find that the size evolution  depends on the stellar mass.  We give
tentative scalings between the power-law index of the $R_e-z$ relation
and  stellar mass.   Future surveys,  such as  the  coming Multi-Cycle
Treasury  programs with  \hst,  will have  the  unique opportunity  to
improve  upon our  scalings and  extend to  both higher  redshifts and
lower stellar masses.

\acknowledgments  Special  thanks   are  due  to  D.~Wittman,  P.~Gee,
C.~Peng, J.~Bosch, S.~Schmidt, and P.~Thorman.  We are grateful to the
men and women who worked tirelessly  for many years to make Wide Field
Camera~3  the  instrument it  is  today,  and  to the  STScI  Director
M.~Mountain for the discretionary  time to make this program possible.
Support for  \hst\ program 11359  was provided by NASA  through grants
GO-11359.0\*.A from  the Space  Telescope Science Institute,  which is
operated by the Association of Universities for Research in Astronomy,
Inc., under NASA contract  NAS 5-26555.  RAW acknowledges support from
NASA  JWST  Interdisciplinary Scientist  grant  NAG5-12469 from  GSFC.
Finally, we are deeply indebted to the brave astronauts of STS-125 for
upgrading and extending \hst\ into the future.

{\it Facilities:} \facility{HST (WFC3)}, \facility{HST (ACS)}

\end{document}